\documentclass[paper,showpacs,preprintnumbers,amsmath,amssymb]{revtex4}
\usepackage{graphicx}% Include figure files
\usepackage{dcolumn}% Align table columns on decimal point
\usepackage{bm}% bold math

\newcommand{\be}{\begin{equation}}
\newcommand{\en}{\end{equation}}
\newcommand{\bea}{\begin{eqnarray}}
\newcommand{\ena}{\end{eqnarray}}
\begin{document}

%\preprint{GACG/04/2005}

\title{ Curvaton reheating in tachyonic inflationary models }

\author{Cuauhtemoc Campuzano}
 \email{ccampuz@mail.ucv.cl}
\affiliation{ Instituto de F\'{\i}sica, Pontificia Universidad
Cat\'{o}lica de Valpara\'{\i}so, Casilla 4059, Valpara\'{\i}so,
Chile.}
\author{Sergio del Campo}
 \email{sdelcamp@ucv.cl}
\affiliation{ Instituto de F\'{\i}sica, Pontificia Universidad
Cat\'{o}lica de Valpara\'{\i}so, Casilla 4059, Valpara\'{\i}so,
Chile.}
\author{Ram\'on Herrera}
\email{rherrera@unab.cl} \affiliation{Departamento de Ciencias
F\'\i sicas, Universidad Andr\'es Bello, Avenida Rep\'ublica 237,
Santiago, Chile.}

\date{\today}% It is always \today, today,
             %  but any date may be explicitly specified

\begin{abstract}
The curvaton reheating in a tachyonic inflationary universe model
with an exponential potential is studied. We have found that the
energy density in the kinetic epoch, has a complicated
dependencies of the scale factor. For different scenarios, the
temperature of reheating is computed. These temperature result to
be analogous to those obtained in the standard case of the
curvaton scenario.
\end{abstract}

% PACS, the Physics and Astronomy
                             % Classification Scheme.
%\keywords{Suggested keywords}%Use showkeys class option if keyword
                              %display desired
\maketitle

Inflationary universe models \cite{guth} have solved many problems
of the Standard Hot Big Bang scenario, for example, the flatness,
the horizon, and the monopole problems, among others. In addition,
its has provided a causal interpretation of the origin of the
observed anisotropy of the cosmic microwave background (CMB)
radiation, and also the distribution of large scale structures. In
the standard inflationary universe models, the acceleration of the
universe is driven by a scalar field $\phi$ (inflaton) with an
specific scalar potential, and the quantum fluctuations associated
to this field generate the density perturbations seeding the
structure formations at late time in the evolution of the
universe.  To date, the accumulating observational data,
especially those coming from the CMB observations of WMAP
satellite \cite{astros} indicate the power spectrum of the
primordial density perturbations is nearly scale-invariant, just
as predicted by the single-field inflation in the context of
``slow-roll'' over.

At the end of inflation the energy density of the universe is
locked up in a combination of kinetic and potential energies of
the scalar field, which drives inflation \cite{Lyth1}. One path to
defrost the universe after inflation is known as reheating
\cite{Kolb1}. Elementary theory of reheating was developed in
\cite{Dolgov} for the new inflationary scenario. During reheating,
most of the matter and radiation of the universe are created
usually via the decay of the scalar field that drives inflation,
while the temperature grows in many orders of magnitude. It is at
this point where the universe coincides with the Big-Bang model.

Of particular interest is a quantity known as the reheating
temperature. The reheating temperature  is associated to the
temperature of the universe when the Big Bang scenario begins,
that is when the radiation epoch begins. In general, this epoch is
generated by the decay of the inflaton field, which leads to a
creation of particles of different kinds.

The stage of oscillations of the scalar field is a essential part
for the standard mechanism of reheating. However, there are some
models where the inflaton potential does not have a minimum and
the scalar field does not oscillate. Here, the standard mechanism
of reheating does not work \cite{Kofman}. These models are known
in the literature like non-oscillating models, or simply NO models
\cite{fengli,Felder1}. The NO models correspond to runaway fields
such as module fields in string theory which are potentially
useful for inflation model-building because they presents flat
directions which survive the famous $\eta$-problem of
inflation\cite{dine}. This problem is related to the fact that
between the inflationary plateau and the quintessential tail there
is a difference of over a hundred orders of magnitude. On the
other hand, an important use of NO models is quintessential
inflation, in which the tail of the potential can be responsible
for the accelerated expansion of the present universe
\cite{Dimopoulus}.

The first mechanism of reheating in this kind of model was the
gravitational particle production \cite{ford}, but this mechanism
is quite inefficient, since it may lead to certain cosmological
problems \cite{ureña,Sami_taq}. An alternative mechanism of
reheating in NO models is the instant preheating, which introduce
an interaction between the scalar field responsible for inflation
an another scalar field \cite{Felder1}.

An alternative mechanism of reheating in NO models is the
introduction of the curvaton field \cite{ref1u}. The decay of the
curvaton field into conventional matter offers an efficient
mechanism of reheating, and does not necessarily introduce an
interaction between the scalar field responsible of inflation and
another scalar field \cite{fengli}. The curvaton field has the
property whose energy density is not diluted during inflation, so
that the curvaton may be responsible for some or all the matter
content of the universe at present.

Implications of string/M-theory to Friedman-Robertson-Walker
cosmological models have recently attracted great attention,
especially those related to brane-antibrane configurations as
spacelike branes. The tachyon field associated with unstable
D-branes, might be responsible for cosmological inflation at early
evolution of the universe, due to tachyon condensation near the
top of the effective scalar potential \cite{sen1,sen2} which could
add some new form of cosmological dark matter at late times
\cite{Sami_taq}. In fact, historically, as was empathized by
Gibbons\cite{gibbons}, if the tachyon condensate starts to roll
down the potential with small initial $\dot{\phi}$, then a
universe dominated by this new form of matter will smoothly evolve
from a phase of accelerated expansion (inflation) to an era
dominated by a non-relativistic fluid, which could contribute to
the dark matter specified above. We should note that during
tachyonic inflation, the slow-roll over condition becomes
$\dot{\phi}^2<2/3$ which is very different from the condition for
non-tachyonic field $\dot{\phi}^2<V(\phi)$. In this way, the
tachyonic field should start rolling with a small value of
$\dot{\phi}$ in order to have a long period of
inflation\cite{Sami_taq}. In this way, the basic field Eqs. for
tachyon inflation become $3H\dot{\phi}+(1/V)(dV/d\phi)\approx 0$
and $3 H^2\approx \kappa_0 V$, where $\phi$ denotes a homogeneous
tachyonic field (with unit $1/m_p$--$1/energy$-- $m_p$ is the
Planck mass, so that $\dot{\phi}$ becomes dimensionless).
$V=V(\phi)$ is the tachyonic potential, $H$ is the Hubble factor
and $\kappa_0=8\pi m_p^{-2}$. We have used units in which
$c=\hbar=1$. Dots mean derivatives with respect to time. These
expressions should be compared with those corresponding to the
standard case, where we have $3H\dot{\phi}+dV/d\phi\approx 0$ and
$3H^2 \approx \kappa_0 V$. Thus, we observe a clear difference in
the scalar field evolution equations, meanwhile the Friedman
equation remains practically the same. Certainly, this
modifications has important consequences, for instance, the
slow-roll over parameters become quite differents \cite{chinos}.

In the following, we explore the curvaton reheating in tachyonic
inflationary models with an exponential potential (i.e. a NO
model). We follow a similar procedure described in
Ref.\cite{ureña}. As the energy density decreases, the tachyonic
field makes a transition into a kinetic energy dominated regime
bringing inflation to the end. Following Liddle and Ure\~na
\cite{ureña}, we considered the evolution of the curvaton field
through three different stages. Firstly, there is a period in
which the tachyonic energy density is the dominant component, i.e,
$\rho_\phi\gg\rho_\sigma$, even though the curvaton field survives
the rapid expansion of the universe. The following stage i.e.,
during the kinetic epoch \cite{refere3}, is that in which the
curvaton mass becomes important. In order to prevent a period of
curvaton-driven inflation, the universe must remain tachyon-driven
until this time. When the effective mass of the curvaton becomes
important, the curvaton field starts to oscillate around at the
minimum of its potential. The energy density, associated to the
curvaton field, starts to evolve as non-relativistic matter. At
the final stage, the curvaton field decays into radiation and then
the standard Big-Bang cosmology is recovered afterwards. In
general, the decay of the curvaton field should occur before
nucleosynthesis happens. Other constraints may arise depending on
the epoch of the decay, which is governed by the decay parameter,
$\Gamma_\sigma$. There are two scenarios to be considered,
depending on whether the curvaton field decays before or after it
becomes the dominant component of the universe.

In the first stage the dynamics of the tachyon field is described
in the slow-roll over approach \cite{Sami_taq}. Nevertheless,
after inflation, the term $V^{-1}\partial V/\partial\phi$ is
negligible compared to the friction term. This epoch is called
`kinetic epoch' or `kination' \cite{refere3}, and we will use the
subscript `k' to label the value of the different quantities at
the beginning of this epoch. The kinetic epoch does not occur
immediately after inflation, may exist a middle epoch where the
tachyonic potential force is negligible respect to the friction
term \cite{Guo}.

The dynamics of the Friedman-Robertson-Walker cosmology for the
tachyonic field in the kinetic regimen, is described by the
equations (see \cite{Guo}):
\begin{equation}
\frac{\ddot{\phi}}{1-\dot{\phi}^2}+3\,H\,\dot{\phi}=0,
\label{key_1}
\end{equation}
and
\begin{equation}
 3\,H^2\,=\kappa_0\rho_{\phi}\label{key_2}.
\end{equation}
The associated energy density of the tachyonic field, $\rho_\phi$,
is given by the expression:

\begin{equation}
\rho_\phi=\frac{V(\phi)}{\sqrt{1-\dot{\phi}^2}}.
\label{rhoi}
\end{equation}

The tachyonic potential $V(\phi)$ is such that satisfies
$V(\phi)\rightarrow 0$ as $\phi\rightarrow\infty$. It has been
argued \cite{Sen1} that the qualitative tachyon dynamics of string
theory can be describe by an exponential potential of the form
\begin{equation}
V(\phi)=V_0e^{-\alpha\sqrt{\kappa_0}\phi},
\end{equation}
where $\alpha$ and $V_0$ are free parameters. In the following we
shall take $\alpha>0$. We should note that $\alpha\sqrt{\kappa_0}$
represents the tachyon mass\cite{fairbain}. An estimation of these
parameters are given in Ref.\cite{Sami_taq}, where $V_0\sim
10^{-10}m_p^4$ and $\alpha\sim 10^{-5}m_p^2$.

From Eq.~(\ref{key_1}) we find a first integral for $\dot{\phi}$
in terms of $a$ given by
\begin{equation}
\dot{\phi}^2=\frac{1}{1+C\,a^6}\,;\,\;\;\;\;\;\;\;
C=\frac{1-\dot{\phi_k}^2}{\dot{\phi_k}^{2}a_k^{6}}\,
>0,\label{firin}
\end{equation}
where $C$ is an integration constant, $\dot{\phi_k}$ and $a_k$
represent values at the beginning of the kinetic epoch for the
time derivative of the tachyonic field and the scale factor,
respectively. A universe dominated by tachyonic field would go
under accelerate expansion if $\dot{\phi}^2<\frac{2}{3}$. The end
of inflation is characterized by the value
$\dot{\phi}_{end}^2=\frac{2}{3}$. The value of $\dot{\phi}$ at the
beginning of the kinetic epoch lies in the range
$1\gtrsim\dot{\phi_k}^2\gtrsim\frac{2}{3}$.

From Eq.(\ref{rhoi}), after substituting the scalar potential
$V(\phi)$ and $\dot{\phi}^2$ from Eq.(\ref{firin}) into
Eq.(\ref{key_2}), and considering that
$\dot{a}=({d\,a}/{d\,\phi})\,\,\dot{\phi}=
(1+Ca^6)^{-1/2}{d\,a}/{d\,\phi}$, we get,

\begin{equation}
V(\phi)=V(\phi(a))=V:=\left[V_0^{1/2}\,e^{-\alpha\sqrt{\kappa_0}\phi_k/2}
-\sqrt{3}\;\frac{\alpha}{2}C^{1/4}I \right]^2, \label{potent}
\end{equation}
where $I$ represents the integral

\begin{equation}
I(a):=I\;:=\;
\int_{a_k}^{a}\frac{a'^{1/2}}{(1+Ca'^6)^{3/4}}\;d\,a'
  =\frac{2}{3}a^{2/3}{}_2F_1\left[\frac{1}{4},\frac{3}{4},\frac{5}{4};-Ca^6\right]
  -\frac{2}{3}a_k^{2/3}{}_2F_1\left[\frac{1}{4},\frac{3}{4},\frac{5}{4};-Ca_k^6\right],
\label{int}
\end{equation}
and ${}_2F_1$ is the  hypergeometric function.

Now from Eq.(\ref{rhoi}) we get an explicit expression for the
tachyonic energy density in terms of the scale factor
\begin{equation}
  \rho_{\phi}=\frac{V}{\sqrt{1-\dot{\phi}^2}}=\frac{\sqrt{1+Ca^6}}{\sqrt{C}a^3}
  \left[V_0^{1/2}\,e^{-\alpha\sqrt{\kappa_0}\phi_k/2}-\sqrt{3}\;
  \frac{\alpha}{2}C^{1/4}I
   \right]^2.\label{Impor}
\end{equation}

Finally, in the kinetic epoch the tachyonic energy density and the
Hubble factor can be written as follows,
\begin{eqnarray}
\rho_{\phi}&=&\rho_{\phi}^{k}\frac{V}{V_{k}}\frac{a_{k}^3}
{\sqrt{Ca_{k}^6+1}}\frac{\sqrt{Ca^6+1}}{a^3}\label{rho},\\
H&=&H_{k}\left(\frac{V}{V_{k}}\right)^{1/2}\frac{a_{k}^{3/2}}
{(Ca_{k}^6+1)^{1/4}}\frac{(Ca^6+1)^{1/4}}{a^{3/2}}  \label{h},
\end{eqnarray}
respectively. Here, $H_{k}^2=\frac{\kappa_0}{3}\rho_\phi^{k}$.
Note the difference for the standard case, where the energy
density is definite by
$\rho_{\phi}^{(std)}=\dot{\phi}^2/2+V(\phi)$, and has a behavior
during the kinetic epoch like stiff matter, i.e.,
$\rho_{\phi}^{(std)}=\rho_\phi^{k}(a_k/a)^6 $ \cite{ureña}. In
this way, the Hubble factor in the standard theory follows the law
$H^{(std)}=H_k(a_k/a)^3$.

We now study the dynamic of the curvaton field, $\sigma$, through
different stages. This permits us to find some constraints of the
parameters, and thus, to have a viable curvaton scenario. We
considered that the curvaton field obeys the Klein-Gordon
equation, and for simplicity, we assume that its scalar potential
is given by
\begin{equation}
U(\sigma)=\frac{m^2\sigma^2}{2}\;,
\end{equation}
where $m$ is the curvaton mass.

First of all, it is assumed that the tachyonic energy density,
$\rho_{\phi}$, is the dominant component when it is compared with
the curvaton energy density, $\rho_\sigma$. In the next stage, the
curvaton field oscillates around the minimum of the effective
potential $U(\sigma)$. Its energy density evolves as a
non-relativistic matter, and during the kinetic epoch the universe
remains tachyonic-dominated. The last stage corresponds to the
decay of the curvaton field into radiation, and then, the standard
Big-Bang cosmology is recovered.

In the inflationary regimen is supposed that the curvaton mass
satisfied the condition $m\ll\,H_f$ and its dynamics is described
in detail in Refs.\cite{dimo,postma,ureña}. During inflation, the
curvaton would roll down its potential until its kinetic energy is
depleted by the exponential expansion and only then, i.e. only
after its kinetic energy is almost vanished, it becomes frozen and
assumes roughly a constant value i.e $\sigma_i\approx \sigma_f$.
Here the subscripts $i$ and $f$ are used to denote the beginning
and the end of inflation, respectively.

The hypothesis is that during the kinetic epoch  the Hubble
parameter decreases so that its value is comparable with the
curvaton mass i.e., $m\simeq H$. From Eq.(\ref{h}), we obtain

\begin{equation}
\frac{m}{H_{k}}=\left(\frac{V_m}{V_{k}}\right)^{1/2}\frac{a_{k}^{3/2}}
{(Ca_{k}^6+1)^{1/4}}\frac{(Ca_m^6+1)^{1/4}}{a_m^{3/2}},\label{mh}
\end{equation}
where the `m' label represents the quantities at the time when
the curvaton mass is of the order of $H$ during the kinetic epoch.

In order to prevent a period of curvaton-driven inflation, the
universe must still be dominated by the tachyonic matter, i.e.
$\rho_{\phi}|_{a_m}=\rho_{\phi}^{(m)}\gg\rho_{\sigma}(\sim\,U(\sigma_f)\simeq\,U(\sigma_i))$
. This inequality allows us to find a constraint on the initial
values of the curvaton field in the inflationary epoch. Hence,
from Eq.(\ref{key_2}), at the moment when $H\simeq m$ we get the
restriction

\begin{equation}
\frac{m^2\sigma_i^2}{2\rho_\phi^{(m)}}=\frac{4\pi}{3}
\frac{m^2\sigma_i^2}{m_p^2m^2}\ll1 \Rightarrow
\sigma_i^2\ll\frac{3}{4\pi}m_p^2 .\label{pot}
\end{equation}

This value is the same to that found in the standard case
\cite{ureña}.

The ratio between the potential energies at the end of inflation
is given by

\begin{equation}
\frac{U_f}{V_f}=\frac{4\pi}{3}\frac{m^2\sigma_i^2}{m_p^2 H_f^2}\ll
\frac{m^2}{H_f^2 }\label{u}.
\end{equation}
Here, we have used for $V_f=(3/8\pi) H_f^2m_p^2$ and
Eq.(\ref{pot}). Thus, the curvaton mass should obey the constraint
in the tachyonic model

\begin{equation}
m\ll H_f, \label{one}
\end{equation}
which gives from Eq.(\ref{u}) that $U_f \ll V_f$. We should note
that the condition given by Eq.(\ref{one}) is inherent to the
nature of the curvaton field, since the reason $m\ll H_f$ is
because only then can be the curvaton superhorizon perturbations
of $\sigma$ be generated during inflation. In this way, the
condition $m\ll H_f$ is a fundamental prerequisite for the
curvaton mechanism.

After the mass of curvaton field becomes important, i.e. $m\simeq
H$, its energy decays like non-relativistic matter in the form

\begin{equation}
\rho_\sigma =
\frac{m^2\sigma_i^2}{2}\frac{a_m^3}{a^3}\label{c_cae}.
\end{equation}

As we have claimed the curvaton decay could be occur in two
different possible scenarios. In the first scenario, when the
curvaton comes to dominates the cosmic expansion (i.e.
$\rho_\sigma>\rho_\phi$), there must be a moment when the
tachyonic  and curvaton energy densities becomes equal. From Eqs.
(\ref{rho}), (\ref{h}) and (\ref{c_cae}) at the time when
$\rho_\sigma=\rho_\phi$, which happens when $a=a_{eq}$, we get

\begin{equation}
\left.\frac{\rho_\sigma}{\rho_\phi}\right|_{a=a_{eq}}=
\frac{4\pi\sigma_i^2 m^2}{3H_{k}^2m_p^2}
\frac{V_{k}a_m^3\sqrt{Ca_{k}^6+1}}
{V_{eq}a_{k}^3\sqrt{Ca_{eq}^6+1}} =\frac{4\pi\sigma_i^2
}{3m_p^2}\frac{V_m}{V_{eq}}
\sqrt{\frac{Ca_m^6+1}{Ca_{eq}^6+1}}=1.\label{equili}
\end{equation}
Now from Eqs.(\ref{h}),(\ref{mh})and(\ref{equili}), we obtain a
relation for the Hubble parameter, $H_{eq}$, in terms of curvaton
parameters and the ratio of the scale factor at different times,
given by:

\begin{equation}
H_{eq}=H_{k}\left(\frac{V_{eq}}{V_{k}}\right)^{1/2}\frac{a_{k}^{3/2}}
{(Ca_{k}^6+1)^{1/4}}\frac{(Ca_{eq}^6+1)^{1/4}}{a_{eq}^{3/2}}=
\sqrt{\frac{4\pi\sigma_i^2}{3m_p^2}}\,\,
\left[\frac{a_m}{a_{eq}}\right]^{3/2}\;m. \label{heq}
\end{equation}
Notice that this result coincides with the one obtained in
standard case.

On the one hand, the decay parameter $\Gamma_\sigma$ is
constrained by nucleosynthesis. For this, it is required that the
curvaton field decays before of nucleosynthesis, which means
$H_{nucl}\sim 10^{-40}m_p < \Gamma_\sigma$. On the other hand, we
also require that the curvaton decay occurs after $\rho_\sigma >
\rho_\phi$, and $\Gamma_\sigma < H_{eq}$ so that we get a
constraint on the decay parameter,

\begin{equation}
10^{-40}m_{p}<\Gamma_{\sigma}<\sqrt{\frac{4\pi\sigma_i^2}{3m_p^2}}
\,\,\left[\frac{a_m}{a_{eq}}\right]^{3/2}\,m. \label{gamm1}
\end{equation}

Until now, it is interesting to give an estimation of the
constraint of the parameters of our model, by using the scalar
perturbation related to the curvaton field. During the time the
fluctuations are inside the horizon, they obey the same
differential equation as the inflaton fluctuations do, from which
we conclude that they acquire the amplitude $\delta\sigma_i\simeq
H_i/2\pi$. Once the fluctuations are outside the horizon, they
obey the same differential equation that the unperturbed curvaton
field does and then we expect that they remain constant during
inflation. The spectrum of the Bardeen parameter $P_\zeta$, whose
observed value is about $2\times 10^{-9}$, allows us to determine
the initial value of the curvaton field in terms of the parameter
$\alpha$. At the time when the decay of the curvaton fields occur,
the Bardeen parameter becomes \cite{ref1u}

\begin{equation}
P_\zeta\simeq \frac{1}{9\pi^2}\frac{H_i^2}{\sigma_i^2}.
\label{pafter}
\end{equation}
The spectrum of fluctuations is automatically gaussian for
$\sigma_i^2\gg H_i^2/4\pi^2$, and is independent of
$\Gamma_\sigma$ \cite{ref1u}. This feature will simplify the
analysis in the space parameter of our models. Moreover, the
spectrum of fluctuations is the same as in the standard scenario.

From Eq.(\ref{pafter}) and by using that $H_i^2=H_f^2(2N+1)$ and
$H_f^2=\alpha^2\,\kappa_0\,/6$ \cite{Sami_taq}, we could relate
the perturbation with the parameters of the model in such way that
we could write

\begin{equation}
\frac{27\pi}{4}\frac{P_{\zeta}}{(2N+1)}\,\sigma_i^2=
\frac{\alpha^2}{m_p^2}. \label{18}
\end{equation}
This expression  allows us to the above equation permit fix the
initial value of the curvaton field in terms of the free parameter
$\alpha$. By using Eq.(\ref{18}), the constraint Eq.(\ref{one})
becomes

\begin{equation}
\frac{m}{m_p}\ll
3\pi\frac{P^{1/2}_\zeta}{(2N+1)^{1/2}}\frac{\sigma_i}{m_p}.\label{21}
\end{equation}

Finally, Eq.(\ref{gamm1}) restricts the value of the decay
parameter $\Gamma_\sigma$, which can be transformed into another
constraint upon $m$ and $\sigma_i$, so that

\begin{equation}
\frac{m}{m_p}\sqrt{\frac{\sigma_i^2}{m_p^2}}\gg\sqrt{\frac{3}{4\pi}}\times
10^{-40},
\end{equation}
where we have used the condition $a_m< a_{eq}$, and
Eq.(\ref{gamm1}).

On the other hand, for the second scenario, the decay of the field
happens before this it dominates the cosmological expansion, that
is, we need that the curvaton field decays before that its energy
density becomes greater than the tachyonic one. Additionally, the
mass is no-negligible so that we could use Eq.(\ref{c_cae}). The
curvaton decays at a time when $\Gamma_\sigma =H$ and then from
Eq.(\ref{h}) we get

\begin{equation}
\frac{\Gamma_\sigma}{H_k}=\left(\frac{V_d}{V_{k}}\right)^{1/2}\frac{a_{k}^{3/2}}
{(Ca_{k}^6+1)^{1/4}}\frac{(Ca_d^6+1)^{1/4}}{a_d^{3/2}},
\label{Gamm}
\end{equation}
where ` d' labels the different quantities at the time when the
curvaton decays, allowing the curvaton field decays after the mass
takes importance, so that $\Gamma_\sigma<m$; and before that the
curvaton field dominates the expansion of the universe, i.e.,
$\Gamma_\sigma>H_{eq}$ (see Eq.(\ref{heq})). Thus,

\begin{equation}
\sqrt{\frac{4\pi\sigma_i^2}{3m_p^2}}\,\,
\left[\frac{a_m}{a_{eq}}\right]^{3/2}m\,<\Gamma_\sigma<m.
\label{gamm2}
\end{equation}

Notice that the range of $\Gamma_\sigma$ is the same that obtained
in the standard case.

Now for the second scenario, the curvaton decays at the time when
$\rho_\sigma<\rho_\phi$. If we defined the $r_d$ parameter as the
ratio between the curvaton and the tachyonic energy densities,
evaluated at $a=a_d$ and for $r_d\ll 1$, the Bardeen parameter is
given by \cite{ref1u,L1L2}

\begin{equation}
P_\zeta\simeq \frac{r_d^2}{36\pi^2}\frac{H_i^2}{\sigma_i^2}.
\label{pbefore}
\end{equation}
With the help of Eq. (\ref{Gamm}) we obtain

\begin{equation}
r_d=\left.\frac{\rho_\sigma}{\rho_\phi}\right|_{a=a_d}=\frac{4\pi\sigma_i^2
m^2}{3H_{k}^2m_p^2} \frac{V_{k}a_m^3\sqrt{Ca_{k}^6+1}}
{V_{d}a_{k}^3\sqrt{Ca_{d}^6+1}}=\frac{4\pi\sigma_i^2
}{3\Gamma_\sigma^2}\frac{m^2}{m_p^2} \frac{a_m^3}{a_d^3}
\label{rd}.
\end{equation}
From Eq.(\ref{gamm2}) we obtain that $r_d<(a_{eq}/a_d)^3$, then
from $r_d\ll 1$ we get $(a_{eq}/a_d)^3\ll 1$. Therefore, the
condition $a_{eq}\ll a_d$ allows us to use expression
(\ref{pbefore}) for the Bardeen parameter.

From expression (\ref{pbefore}) and (\ref{rd}) we could write \be
\frac{\sigma_i^2}{m_p^2}=\frac{81}{4}\frac{m_p^2}{m^4}
\left(\frac{a_d}{a_m}\right)^6\frac{P_\zeta}{(2N+1)}
\frac{\Gamma_\sigma^4}{H_f^2}, \en and thus the expression
(\ref{gamm2}) becomes

\begin{equation}
\sqrt{27\pi}\frac{a_d^3}{a_m^{3/2}a_{eq}^{3/2}}\frac{P_\zeta^{1/2}}
{(2N+1)^{1/2}}\frac{m_p}{m^2
H_f}\Gamma_\sigma^2<\frac{\Gamma_\sigma}{m}<1.\label{newphys}
\end{equation}

Even though the study of  gravitational waves was developed in
Ref.\cite{Kolb} for the tachyonic model, it is interesting to give
an estimation of the constraint on the curvaton mass, using this
type of tensorial perturbation. Under the approximation give in
Ref.\cite{Hwang}, the corresponding gravitational wave amplitude
in the tachyonic model may be written as
$$
h_{GW}^2\simeq\,C_1\,\frac{V_i}{m_p^4},
$$
where the constant $C_1\approx\,10^{-3}$. Now, using that
$V_i=V_f(2N+1)$, we obtain
\begin{equation}
h_{GW}^2\simeq\,3C_1\,(2N+1)\frac{H_f^2}{8\pi m_p^2}.\label{gw}
\end{equation}
In this way, from Eqs.(\ref{one}) and (\ref{gw}) we get that
\begin{equation}
m^2\ll\,8\pi m_p^2\frac{h_{GW}^2}{3C_1(2N+1)}.
\end{equation}

If we consider that $h_{GW}$ of the order of $10^{-5}$ and if take
the number of e-fold to be $N\simeq 70$, (but in context of the
curvaton may be much lower that this value, let say 45 or so,
since the inflationary scale can be lower) we find that the above
equation gives the following upper limit for the curvaton mass

\begin{equation}
m \ll 10^{-4}m_p\sim 10^{15} GeV, \label{29}
\end{equation}
which coincides with the limit reported in Ref.\cite{Dimo}. We
note that in this model we have $V''=3\alpha^2 H_f^2$, where the
prime denotes differentiation with respect to the tachyonic field
$\phi$, and if $V''> H_f^2 m_p^4$ the curvature perturbations can
becomes too large compare to the COBE observations \cite{Dimo}, so
that the inflationary scale cannot be much larger than the scale
of grand unification ( inflation does not produce perturbations if
$\alpha>m_p^2/\sqrt{3}$). This means that $H_{f} \leq 10^{13}GeV$
\cite{Kofman}. Hence, the bounds in Eq.(\ref{29}) is redundant
unless the inflaton does not produce any curvature perturbations.

In order to give an estimation of the gravitational wave, we move
to the kinetic epoch in which the energy density of gravitational
waves evolves as in Refs.\cite{Dimopoulus,referee4}
\begin{equation}
\frac{\rho_g}{\rho_\phi}\sim h_{GW}^2 \left(\frac{V_k}{V}
\frac{a_k}{a}\right)\sqrt{\frac{Ca_k^6+1}{Ca^6+1}}\label{reata},
\end{equation}
 where we have used the same approximation than that used in Ref.
\cite{Hwang}.

On the other hand,  when the curvaton field decays, i.e.
($\Gamma_\sigma =m$) it produces radiation which decays as
$1/a^4$. Then we may write
\begin{equation}
\rho_r^{(\sigma)}=\frac{m^2\sigma_i^2}{2}\frac{a_m^3}{a_d^3}
\frac{a_d^4}{a^4}.
\end{equation}

If the radiation produced from the curvaton scalar field is equal
to the tachyonic density i.e. $\rho_\sigma^{(r)}=\rho_\phi$, at
the time in which $a=a_{eq}$, then we could keep the gravitational
waves stable, so that

\begin{equation}
\left.\frac{\rho_r^{(\sigma)}}{\rho_\phi}\right|_{a=a_{eq}}=\frac{4\pi
m^2}{3
m_p^2}\frac{\sigma_i^2}{H_k^2}\frac{V_k}{V_{eq}}\left(\frac{a_m}{a_k}\right)^3
\left(\frac{a_d}{a_{eq}}\right)\sqrt{\frac{Ca_k^6+1}{Ca_{eq}^6+1}}=1,
\end{equation}
and used Eqs. (\ref{mh}), (\ref{Gamm}) and (\ref{reata}), we
obtain a constrains during the kinetic epoch given by

\begin{equation}
\frac{m\sigma_i}{m_p}\gg
h_{_{GW}}\;H_k\left(\frac{\Gamma_\sigma}{H_k}\right)^{1/3}\left(\frac{V_k}{V_d}\right)^{1/6}
\left(\frac{a_k}{a_m}\right)^{3/2}\left(\frac{Ca_k^6+1}{Ca_d^6+1}\right)^{1/12},\label{36}
\end{equation}
where we have used $\rho_g/\rho_r\ll 1$ at the time in which
$a=a_{eq}$.

We note that from Eqs.(\ref{pot}), (\ref{29}) and (\ref{36}) we
obtain a bound for the $m$ parameter, i.e.

\begin{equation}
\sqrt{\frac{4\pi}{3}}h_{_{GW}}\;H_k\left(\frac{\Gamma_\sigma}{H_k}\right)^{1/3}\left(\frac{V_k}{V_d}\right)^{1/6}
\left(\frac{a_k}{a_m}\right)^{3/2}\left(\frac{Ca_k^6+1}{Ca_d^6+1}\right)^{1/12}\ll
m \ll 10^{-4}m_p\label{lachi}.
\end{equation}
It is interesting to note that in this case we have obtained a
bound from bellow for the $m$ curvaton mass.

In the first scenario, our computes allow to get the reheating
temperature as hight as $10^{-9}m_p$, since the decay parameter
$\Gamma_{\sigma}\propto\,T_{rh}^2/m_p$, where $T_{rh}$ represents
the reheating  temperature. Here, we have used Eqs, (\ref{gamm1}),
(\ref{18}) and (\ref{29}), $a_m/a_{eq}\sim\,10^{-1}$ and
$\alpha\sim\,10^{-5} m_p^2$. We should compare this bound with the
bound coming from gravitino over-production, which gives
$T_{rh}\leq 10^{-10}m_p$ \cite{rtref}.

In the second scenario from the Eqs. (\ref{gamm2}) and (\ref{29}),
we could estimate the reheating temperature to be of the order of
$\sim 10^{-3}m_p$ as an upper limit.

As it was reported in Ref.\cite{Sami_taq} at the end of inflation
$\rho_\phi$ at best could scales as $a^{-3}$, it is valid
irrespectively of the form of the tachyonic potential provides it
satisfies $V(\phi)\rightarrow 0$ as $\phi\rightarrow \infty$.
However, this is not in general since in our particular case, we
have found that it is possible to get a more complex expression
for the dependence of $\rho_\phi$ in terms of the scale factor.
This could be seen from Eq. (\ref{Impor}).

On the other hand, the shape of the tachyon condensate effective
potential depends on the system under consideration. In bosonic
string theory, for instance, this potential has a maximum, $V =
V_0$, at $\phi = 0$, where $V_0$ is the tension of some unstable
bosonic D-brane. A local minimum, $V = 0$, generically at
$\phi\rightarrow\infty$, corresponding to a metastable closed
bosonic string vacuum, and a runaway behavior for negative $\phi$.
An exact classical potential (i.e. exact to all orders in
$\alpha'$, but only at tree level in $g_s$) encompassing these
properties has been considered \cite{belga},
\begin{equation}
V(\phi)=V_0(1+\phi/\phi_0)\exp(-\phi/\phi_0),
\end{equation}
where the parameters $V_0$ and $\phi_0$ in terms of the string
length $l_s$ and the open string coupling constant $g_s$, are
given by
\begin{equation}
 V_0 = \frac{v_0}{ g_s l_s^4(2\pi)^3} , \phi_0 =\frac{1}
 {\alpha \sqrt{\kappa_0}}= \tau_0 l_s,
\end{equation}
with $v_0$ and $\tau_0$ dimensionless parameters, such that
$V_0/v_0$ is the tension of a D3-brane and $\tau_0 l_s$ is the
inverse tachyon mass \cite{fairbain}. The gravitational coupling
in 4 dimensions is given in terms of the stringy parameters by

\begin{equation}
\kappa_0\equiv 8\pi G_N=\frac{8\pi}{m_p^2}= \pi g_s^2
l_s^2\left(\frac{l_s}{R}\right)^6=\frac{ g_s^2 l_s^2}{v}\; ;\;
v=\frac{1}{\pi}\left(\frac{R}{l_s}\right)^6.
\end{equation}
Here, $R$ is the compactification radius of the compact 6
dimensional manifold, taken to be a 6-torus. For the D = 4
effective theory to be applicable, one usually requires that $R
\gg l_s$ i.e. $v\gg1$.

From Eq.(\ref{newphys}) we obtain that

\begin{equation}
36\pi\frac{a_d^3}{(a_m a_{eq})^{3/2}}\frac{P^{1/2}_\zeta
}{(2N+1)^{1/2}}\frac{\Gamma_\sigma^2}{m^2}< \frac{g_s}{\tau_0
v^{1/2}}.\label{new}
\end{equation}

From Eq.(\ref{new}) and taking $N=50$, $P_\zeta\sim 10^{-5}$, and
$\tau_0=1$, we find a constraint (from the reheating scenario) for
the parameter $g_s$ coming from string theory, which is given by

\begin{equation}
g_s^2> 10^{-4}\frac{a_d^6}{(a_m
a_{eq})^{3}}\frac{\Gamma^4_\sigma}{m^4}\;v,
\end{equation}
the above expression give to us a lower bound for the string
coupling constant. From the amplitude of gravitational waves
produced during inflation the upper bound is $g_s^2\leq 10^{-9} v$
\cite{Kofman}. In this way, we have the following constraint for
$g_s^2/v$
\begin{equation}
 10^{-4}\frac{a_d^6}{(a_m a_{eq})^{3}}\frac{\Gamma^4_\sigma}{m^4}
 < \frac{g_s^2}{v}\leq 10^{-9} .
\end{equation}
Summarizing, we have describe curvaton reheating in tachyonic
inflationary model in which we have considered two cases. In the
first case the curvaton dominates the universe before it decay.
Our results are specified by Eqs.(\ref{gamm1}) and (\ref{21}), and
we see that they are identical with the standard curvaton scenario
\cite{ureña}. In the second case where the curvaton decaying
before domination, we have arrived to Eq.(\ref{newphys}), which
represents one of the most important constraint by using the
curvaton approach.

In conclusion, we have introduced the curvaton mechanism into NO
inflationary tachyonic model  as another possible solution to the
problem of reheating, where there is not need to introduce an
interaction between the tachyonic and some auxiliary scalar field.

\begin{acknowledgments}
CC was supported by MINISTERIO DE EDUCACION through MECESUP Grants
FSM 0204. SdC was supported by COMISION NACIONAL DE CIENCIAS Y
TECNOLOGIA through FONDECYT grants N$^0$ 1030469, N$^0$1040624 and
N$^0$1051086. Also from UCV-DGIP N$^0$ 123.764 and from
Direcci\'on de Investigaci\'on UFRO N$^0$ 120228. RH was supported
by UNAB  Grants DI 28-05/R.

\end{acknowledgments}

%\\\\\\\\\\\\\\\\\\\\\\\\\\\\\\\\\\\\\\\\\\\\\\\\\\\\\\\\\\\\\\\\\\\\\\\


\begin{thebibliography}{99}


\bibitem{guth} A. Guth, Phys. Rev. D {\bf 23}, 347 (1981); A.
Albrecht and P. J. Steinhardt, Phys. Rev. Lett. {\bf 48}, 1220
(1982); A. Linde, Phys. Lett. B {\bf 108}, 389 (1982).

\bibitem{astros} C. L. Bennet, Astrophys. J. Suppl. {\bf 148}, 1 (2003); D. N.
Spergel, Astrophys. J. Suppl. {\bf 148}, 175 (2003); H.V. Peiris,
Astrophys. J. Suppl. {\bf 148}, 213 (2003).

\bibitem{Lyth1}D. H. Lyth and A. Riotto, Phys. Rept. B{\bf 314}, 1
(1999).

\bibitem{Kolb1} E. W. Kolb and M. S. Turner, {\it The Early
Universe}, (Addison-Wesley, Menlo Park, Ca., 1990).

\bibitem{Dolgov} A. D. Golgov and A. Linde, Phys. Lett. B{\bf 116},
329 (1982); L. F. Abbott, E. Fahri and M. Wise, Phys. Lett. B{\bf
117}, 29 (1982).

\bibitem{Kofman} L. Kofman and A. Linde, JHEP {\bf 0207}, 004 (2002).


\bibitem{Felder1} G. Felder, L. Kofman and A. Linde, Phys. Rev. D{\bf 60},
103505 (1999).

\bibitem{fengli}B. Feng and M. Li, Phys. Lett. B {\bf 564} 169
(2003).

\bibitem{dine} M. Dine, L. Randall and S. Thomas, Nucl. Phys. B{\bf 458}, 291
(1996); Phys. Rev. Lett. {\bf 75}, 398 (1995).

\bibitem{Dimopoulus} K. Dimopoulus, Phys. Rev. D {\bf 68}, 123506
(2003).

\bibitem{ford} L. H. Ford, Phys. Rev. D{\bf 35}, 2955
(1987).

\bibitem{ureña}A. R. Liddle and L. A. Ure\~na-L\'opez,
Phys. Rev. D{\bf 68}, 043517 (2003).

\bibitem{Sami_taq}M. Sami, P. Chingangbam and T. Qureshi,
Phys. Rev. D{\bf 66}, 043530 (2002).

\bibitem{ref1u}D. H. Lyth and D. Wands, Phys. Lett. B{\bf 524}, 5
(2002); S. Mollerach, Phys. Rev. D {\bf 42}, 313 (1990).










\bibitem{sen1} A. Sen, JHEP  {\bf 04} 048 (2002).

\bibitem{sen2} A. Sen,  Mod. Phys. Lett. A {\bf 17} 1797 (2002).

\bibitem{gibbons} G. W. Gibbons, Phys. Lett. B {\bf 537} 1 (2002).

\bibitem{chinos} X. Li, D. Liu and J. Hou,  J. Shangai Normal Univ.
(Natural Science), {\bf 33(4)} 29 (2004).

\bibitem{refere3} M. Joyce and T. Prokopec, Phys. Rev. D{\bf 57}, 6022 (1998).

\bibitem{Guo} Z. Guo, Y. Piao, R. Cai and Y. Zhang, Phys. Rev. D {\bf 68}, 043508 (2003).

\bibitem{Sen1} A. Sen,  Mod. Phys. Lett. A {\bf 17} 1797 (2002) .

\bibitem{fairbain} M. Fairbairn and M. H. G. Tytgat, Phys. Lett. B {\bf 546} 1 (2002).

\bibitem{dimo}  K. Dimopoulos, G. Lazarides, D. H. Lyth and R. Ruiz de Austri,
Phys. Rev. D {\bf 68} 123515 (2003).

\bibitem{postma}  M. Postma,  Phys. Rev. D {\bf 67}, 063518 (2003).

\bibitem{L1L2}D. H. Lyth, C. Ungarelli and  D. Wands, Phys. Rev. D {\bf
67}, 023503 (2003).

\bibitem{Kolb} J. Linsey, A. Liddle, E. Kolb and E. Copeland,
Rev. Mod. Phys.  {\bf 69}, 373 (1997).


\bibitem{Hwang} J. Hwang and H. Noh, Phys. Rev. D  {\bf 66},
084009 (2002).

\bibitem{Dimo}  K. Dimopoulos and D. H. Lyth,  Phys. Rev. D {\bf 69}, 123509 (2004).


\bibitem{referee4} V. Sahni, M. Sami and T. Souradeep, Phys. Rev. D{\bf 65},
023518 (2002); M. Giovannini, Class. Quant. Grav. {\bf 16}, 2905
(1999); Phys. Rev. D{\bf 60}, 123511 (1999).

\bibitem{rtref} J. R. Ellis, J. S. Hagelin, D. V. Nanopoulus, K.
A. Olive and M. Srednicki, Nucl. Phys. B {\bf 238}, 453 (1984); M.
Kawasaki and T. Moroi, Prog. Theor. Phys. {\bf 93}, 879 (1995).


\bibitem{belga} A. A. Gerasimov and S. L. Shatashvili, JHEP {\bf 10}, 034 (2000);
D. Kutasov, M. Marino and G. W. Moore, JHEP {\bf 10},
 045 (2000) .


































\end{thebibliography}
\end{document}